\documentclass[12pt]{article}
\usepackage{amsmath,amssymb,amsfonts,dsfont,geometry,xcolor,authblk}
\numberwithin{equation}{section}
\usepackage[linktoc=all]{hyperref}

\newcommand{\Spin}{\mathrm{Spin}} 
\newcommand{\U}{\mathrm{U}}
\newcommand{\SL}{\mathrm{SL}}  
\newcommand{\SU}{\mathrm{SU}}

\newcommand{\USp}{\mathrm{USp}}  
\newcommand{\CP}{\mathbb{CP}}
 
\newcommand{\tr}{\mathrm{tr}}

\def\red#1{{\color{red}{#1}}}

\def\nn{\nonumber}
\def\bsm{\begin{smallmatrix}}
\def\esm{\end{smallmatrix}}
\def\bpm{\begin{pmatrix}}
\def\epm{\end{pmatrix}}
\def\til{\widetilde}
\def\hat{\widehat}

\def\C{\mathbb{C}} 
\def\P{\mathbb{P}} 
 
\def\Z{\mathbb{Z}}
\def\cA{\mathcal{A}}

\def\cN{\mathcal{N}}
\def\cO{\mathcal{O}}

\def\yt{{\til y}}
\def\zt{{\til z}}

\def\a{{\alpha}}

\def\g{{\gamma}}
\def\G{{\Gamma}}
\def\d{{\delta}}
\def\e{{\epsilon}}

\def\l{{\lambda}}
\def\m{{\mu}}

\def\x{{\xi}}

\def\s{{\sigma}}
\def\t{{\tau}}

\def\om{{\omega}}

\title{Inozemtsev System as Seiberg-Witten Integrable System}
\author[1]{Philip C. Argyres}
\author[2]{Oleg Chalykh}
\author[3]{Yongchao L\"u}
\affil[1]{\small\it  
	 Physics Department, 
	 University of Cincinnati,
	 PO Box 210011, 
	 Cincinnati OH 45221, US}
 \affil[ ]{{\it e-mail}: philip.argyres@gmail.com}
\affil[2]{\small\it  
	School of Mathematics, 
	University of Leeds, 
	Leeds, LS2 9JT, UK }
\affil[ ]{{\it e-mail}: o.chalykh@leeds.ac.uk}
\affil[3]{\small\it
	Department of Physics and Astronomy, 
	Uppsala university,
	Box 516,
	SE-75120 Uppsala,
	Sweden}
\affil[ ]{{\it e-mail}: yongchao.lu@physics.uu.se}
\date{}

\begin{document}
\noindent
\hspace{\fill} UUITP-01/21\\

\begingroup
\let\newpage\relax
\maketitle
\endgroup

\begin{abstract}	
In this work we establish that the Inozemtsev system is the Seiberg-Witten integrable system encoding the Coulomb branch physics of 4d $\mathcal{N}=2$ $\USp(2N)$ gauge theory with four fundamental and (for $N \geq 2$) one antisymmetric tensor hypermultiplets. 
We describe the transformation from the spectral curves and canonical one-forms of the Inozemtsev system in the $N=1$ and $N=2$ cases to the Seiberg-Witten curves and differentials explicitly, along with the explicit matching of the modulus of the elliptic curve of spectral parameters to the gauge coupling of the field theory, and of the couplings of the Inozemtsev system to the field theory mass parameters.
This result is a particular instance of a more general correspondence between crystallographic elliptic Calogero-Moser systems with Seiberg-Witten integrable systems, which will be explored in future work.
\end{abstract}

\newpage

\tableofcontents

\section{Introduction and summary}

Since the dawn of Seiberg-Witten era \cite{SWi, SWii}, it has been recognized \cite{GKMMM95} that there is close connection between 4d $\cN=2$ systems and completely integrable Hamiltonian systems. In particular, Donagi and Witten \cite{DW95} explained that for each 4d $\cN=2$ supersymmetric field theory there exists a complex integrable systems encoding its Coulomb branch physics. Following \cite{Donagi97} we will call such a complex integrable system a Seiberg-Witten integrable system. 

There are no known systematic ways to identify the Seiberg-Witten integrable system for a given 4d $\cN=2$ theory.  Nevertheless, there have been two main effective approaches in this regard.  In the first approach, one tries to match known many-body or spin chain integrable systems with particular 4d $\cN=2$ theories.  There are several notable examples along this line. For instance, 4d $\cN=2$ pure YM theory with simple gauge algebra $\mathrm{G}$ corresponds \cite{Martinec95_1} to the twisted affine Toda chain of type $(\hat{\mathrm{G}}^{(1)})^\vee$, where $(\hat{\mathrm{G}}^{(1)})^\vee$ is the Langlands dual of the untwisted affine Kac-Moody algebra $\hat{\mathrm{G}}^{(1)}$. Another example \cite{Martinec95_2, DHP97} is the elliptic Calogero Moser system of $A_{N-1}$ type which describes the Seiberg-Witten solution of 4d $\cN=2^\ast$ theories with gauge group $\SU(N)$ or $\U(N)$; this type of matching has been generalized to arbitrary simple gauge groups (with $G_2$ as a potential exception) \cite{DHP98Lax}.  It is also proposed \cite{GMMM96, GGM97} that the inhomogeneous $\mathfrak{sl}_2$ XXX spin chain provides solutions to 4d $\cN=2$ $\SU(N_c)$ gauge theories with $N_f \leq 2 N_c$ fundamental hypermultiplets.  See the survey \cite{GM00} for these and further connections.

A second approach identifies Seiberg-Witten integrable systems for a large class of 4d $\cN=2$ supersymmetric field theories as Hitchin systems on Riemann surfaces with tame/wild ramified punctures.  This class of 4d $\cN=2$ supersymmetric field theories are known as class-S theories \cite{Gaiotto09}.  A precursor to this approach is the M-theory solution to certain 4d $\cN=2$ quiver gauge theories engineered with D4-NS5-D6 brane systems \cite{Witten97}.  

These two approaches --- matching to known integrable systems or to Hitchin systems --- have some overlap.
For instance, it is known that the elliptic Calogero Moser system of type $A_{N-1}$ can be interpreted as the $\SU(N)$ Hitchin system on a torus with a puncture \cite{GN95}.  However, for a majority of Hitchin systems there are no explicitly known many-body or spin chain integrable systems. 

In this and upcoming work \cite{CL_ADE}, we will follow the line of the first approach to identify the Seiberg-Witten systems for several series of 4d $\cN=2$ superconformal field theories which all admit F-theory constructions.  A common feature shared by those theories is that their Coulomb branch chiral rings are given by the rings of symmetric polynomials with respect to certain complex reflection groups \cite{CC18}.\footnote{We refer the reader to the appendix in \cite {TZ19} for a nice account of complex reflection groups aimed at physicists.}  On general grounds all the relevant complex reflection groups also need to satisfy various physical constraints including Dirac quantization and electric-magnetic duality which implies the relevant complex reflection groups must be crystallographic --- which means that there exists an invariant full-rank lattice preserved by the complex reflection group.  All such crystallographic groups have been classified \cite{Popov, Goryunov}.

Generalizations of elliptic Calogero-Moser systems --- known as crystallographic elliptic Calogero-Moser systems --- have been constructed for all crystallographic complex reflection groups \cite{EFMV}. Our proposal is that these are candidates for Seiberg-Witten geometries. A nice feature of these integrable systems is that their full set of parameters  matches the mass deformations of classes of $4d$ $\cN=2$ quantum field theories.
For instance, we identify the elliptic Calogero-Moser systems attached to the crystallographic complex reflection groups of type $G(m, 1, N)$ with $m=2, 3, 4, 6$ as Seiberg-Witten integrable systems for $4d$ $\cN=2$ rank $N$ $D_4$ and $E_6$, $E_7$, $E_8$ theories \cite{MN1, MN2, AT07}.  Those theories belong to the the category of class-S theories, therefore their Seiberg-Witten integrable systems admit Hitchin system construction \cite{NX09, BBT09, GR12}.  

In this paper we will focus on the $G(2, 1, N)$ case, which are also known as the Inozemtsev system \cite{Ino89}, which corresponds to $4d$ $\cN=2$ $\USp(2N)$ gauge theory with one antisymmetric and four fundamental hypermultiplets.  Since $G(2, 1, N)$ is the complexification of the Weyl group $W(B_N) \equiv W(C_N)$ and depends on an elliptic modulus, it is natural to guess that it describes the Coulomb branch of a superconformal gauge theory with $\USp(2N)$ or $\Spin(2N{+}1)$ gauge group. What is surprising is that, on the one hand, the Inozemtsev system has no direct Lie-algebraic interpretation, and on the other hand the Inozemtsev systems has the right pattern of couplings to match exactly with a single class of 4d $\mathcal{N}=2$ gauge theories, namely, the $\USp(2N)$ superconformal theories with one antisymmetric and $N_f=4$ fundamental hypermultiplets.

Since the $\USp(2N)$ $N_f=4$ theory admits class-S description, the Inozemtsev system should be equivalent to an $\SU(2N)$ Hitchin system on the orbicurve $T^2 / \Z_2$, and we offer such an interpretation. Furthermore, the Seiberg-Witten solutions for the particular $\USp(2N)$ gauge theories are given in explicit form via an M5 brane construction in \cite{AMP02}.  The equivalence of the Seiberg-Witten solutions with the Inozemtsev system is not at all obvious.  In this work we check their equivalence for the rank $N = 1, 2$ cases.  We find that we need to modify some choices made in \cite{AMP02} in the M5 brane construction of the Seiberg-Witten curve in order to achieve an algebraically transparent matching to the integrable system.

Our recognition of the Inozemtsev system as a Seiberg-Witten integrable system has some independent interest.  Specifically, one may be able to utilize the gauge theory description to extract exactly solvable observables by various powerful techniques including semi-classical methods, supersymmetric localization, the gauge-Bethe correspondence, and the AGT correspondence, and relate them to the Inozemtsev system.

This paper is organized as follows. In section \ref{section:Ino} we discuss various aspects of Inozemtsev system, and introduce the Lax representation following \cite{Tak99,Chalykh_2019}.  Among other things, we give an interpretation of the Inozemtsev system as a Hitchin system on the four-punctured sphere.  In section \ref{section:D4}, after recalling some general properties of the series of $\USp(2N)$ $N_f=4$ theories, we describe the realization of their Coulomb branch physics in terms of M5 brane curves.  In section \ref{section: BCtoM5_1} we describe the transformation from the spectral curves and canonical one-form of the Inozemtsev system in the $N=1$ and $N=2$ cases to the Seiberg-Witten curves and differentials explicitly, along with the variable and parameter matching.  We include an appendix which summarizes some relevant elliptic functions and identities and outlines the derivation of the $N=2$ spectral curve of the Inozemtsev system.

\section{Inozemtsev system}
\label{section:Ino}

\subsection{Hamiltonian description}

The Inozemtsev system, also known as the Calogero--Moser--Sutherland system of $BC_N$-type, is described by the Hamiltonian \cite{Ino89}: 
\begin{equation}\label{ih}
h_2 = \sum_{j=1}^N (p_j^2 - u(q_j))-2g^2\sum_{j< k}^N \left( \wp(q_j - q_k ) + \wp(q_j + q_k )\right)\,, 
\end{equation}
where $\wp(q)$ is the Weierstrass $\wp$-function with periods $1, \t$ and 
\begin{equation}\label{u}
u(q) = \sum_{r=0}^3 g_r^2 \wp(q + \om_r)\,,\qquad 
(\om_0,  \om_1, \om_2, \om_3 ) = 
\left( 0, \frac12, \frac{1 + \t}{2}, \frac\t2 \right)\,.
\end{equation}
Here $(p_i, q_i)$, $i=1, \dots, N$ represent the momenta and positions of $N$ interacting particles on the line, subject to an external field with potential $-u(q)$. Note that we have four coupling constants $g_{0, 1, 2, 3}$ in the $N=1$ case and one additional coupling constant $g$ in the $N \geq 2$ cases. It is customary to assume, in the repulsive regime, that the couplings $g^2$ and $g_r^2$ are real negative.  For our purposes, however, this is not important, as we consider this system on the complexified phase space $\C^{2N}$ with the standard (holomorphic) symplectic structure. As such, it has the underlying symmetry associated with the {\it complex crystallographic group} generated by the translations $q_j\mapsto q_j+1$, $q_j\mapsto  q_j+\t$ together with the arbitrary permutations and sign changes of $q_j$. This corresponds to the group $[G(2,1,N)]^\tau_1$ in the classification \cite{Popov}. 

The Inozemtsev system is known to be completely integrable in Liouville's sense, which means that it admits $N$ independent Poisson-commuting Hamiltonians $h_2, h_4, \dots, h_{2N}$. The higher Hamiltonians are of the form $h_4=\sum_{i<j}p_i^2p_j^2+\ldots$, $h_6=\sum_{i<j<k}p_i^2p_j^2p_k^2+\ldots$, etc., up to lower degree terms. Explicit expressions for $h_{2k}$ are available for the quantum case \cite{OOS1994} from which the classical Hamiltonians are easily obtained.  For instance, in the $N=2$ case the quartic Hamiltonian can be taken as
\begin{align}\label{h4}
h_4 
& = \Big( p_1 p_2 +  g^2\wp(q_1-q_2) - g^2\wp(q_{1}+q_{2}) \Big)^2 \nn\\
& \quad \mbox{}- u(q_1) p_2^2  - u(q_2) p_1^2  +  u(q_1) u(q_2) 
\nn\\
& \quad \mbox{}+\left(u(q_1)+u(q_2)\right)
\left(g^2\wp(q_1-q_2)+g^2\wp(q_1+q_2)\right)
\nn\\
& \quad \mbox{}-2g^2\sum ^{3}_{i=0} g^{2}_{i}\wp ( q_{1} +\omega _{i}) \wp ( q_{2} +\omega _{i}) \,.
\end{align}

\subsection{Lax matrix}

As another manifestation of the integrability of the model \eqref{ih}, it admits a Lax representation, i.e., a pair of matrix-valued functions $L, A\,:\, \C^{2N}\to \mathrm{Mat}(2N, \C)$ such that the Hamiltonian dynamics takes the form $\frac{d}{dt} L=[L, A]$. An immediate corollary is that the quantities $\tr (L^k)$, as well as the eigenvalues of $L$, are constants of motion, which means that $L$ remains {\it isospectral} for all $t$. Originally, Inozemtsev constructed in \cite{Ino89} a Lax pair of size $3N\times 3N$ (see also \cite{Zotov}); other Lax pairs of smaller size have since been found \cite{DHP98Lax, Tak99}.  We will use the Lax matrix of size $2N\times 2N$ from \cite{Tak99}. To write it down, we need the functions $\s_\alpha(x)$ and $v_\alpha(x) := \sum_{r=0}^{3} g_r \s_{2 \alpha}^r(x)$ whose definition and basic properties are given in the Appendix. 
We have: 
\begin{align}\label{lm1}
L & = \sum_{i=1}^N \left(p_iE_{i,i}-p_iE_{i+N, i+N}+v_\alpha(q_i)E_{i, i+N}+v_\alpha(-q_i)E_{i+N,i}\right)\\
& + g\sum_{i\ne j}^N \left(\sigma_{\alpha}(q_{ij})E_{i,j}+\sigma_{\alpha}(q_{ij}^+)E_{i, j+N}+\sigma_{\alpha}(-q_{ij}^+)E_{i+N, j}+\sigma_{\alpha}(-q_{ij})E_{i+N, j+N}\right)\,, \nn
\end{align}
where $E_{i,j}$ are the elementary matrices, and $q_{ij}$, $q_{ij}^+$ are the shorthand notation for $q_i-q_j$ and $q_i+q_j$, respectively. This Lax matrix $L$ contains an auxiliary parameter $\alpha$, usually referred to as the {\it spectral parameter}, so we may write $L(\alpha)$ to emphasize this dependence. We remark that the above expression for $L$ follows closely \cite[(5.15)]{Chalykh_2019}. It corresponds, in a different notation, to (3.37) and (3.39) in \cite{Tak99}.

As a function of $\alpha$, the Lax matrix $L$  has the following important properties.

\noindent{\bf 1.} Periodicity:
\begin{equation}\label{tran}
L(\alpha+1)=L(\alpha)\,,\quad L(\alpha+\tau)=CL(\alpha)C^{-1}\,,
\end{equation}
where 
$C= \begin{bmatrix} D & 0 \\ 0 & D^{-1} \end{bmatrix}$ with $D = \mathrm{diag}(e^{2\pi i q_1}, \dots, e^{2\pi iq_N})$.

\noindent{\bf 2.} Symmetry:
\begin{equation}\label{sym}
L(-\alpha)=-ML(\alpha)M^{-1}\,, \quad\text{where}\  M = \begin{bmatrix}
0 & \mathrm{I}_N  \\ \mathrm{I}_N & 0 \end{bmatrix}.
\end{equation}    

\noindent{\bf 3.} $L$ has simple poles at the half-periods: $L\sim L_i(\a-\om_i)^{-1}+\mathrm{O}(1)$ near $\a=\om_i$. The residues $L_i$ are 
	\begin{align}\label{res1}
	L_i&= - g_i^\vee\begin{bmatrix}
	0 & \mathrm{I}_N  \\
	\mathrm{I}_N & 0 
	\end{bmatrix}\quad (i=1,2,3)\,,\\
\label{res2}
	L_0&=(g-g_0^\vee) \begin{bmatrix}
	0 & \mathrm{I}_N  \\
	\mathrm{I}_N & 0 
	\end{bmatrix} - gT\,,
	\end{align}
where $T$ is the $2N\times 2N$ matrix with $0$'s along the main diagonal and $1$'s elsewhere, and $g_i^\vee$ are the dual parameters,
\begin{equation}\label{ghat} 
	\bpm g^\vee_0\\ g^\vee_1\\ g^\vee_2\\ g^\vee_3\\ \epm = 
	\frac12 \left( {
		\begin{array}{rrrr}
		1 & 1 & 1& 1\\
		1 & 1 & -1 & -1\\
		1 & -1 & 1 & -1\\
		1 & -1 & -1 & 1\\
		\end{array}
	} \right)
	\bpm g_0\\ g_1\\ g_2\\ g_3\\ \epm.
	\end{equation}
Note that the residues $L_i$ are semi-simple (diagonalizable), with
\begin{align}\label{resi1}
L_i&\sim \mathrm{diag} \Big(
\underbrace{-g_i^\vee,\dots, -g_i^\vee}_\text{$N$ times}, \ 
\underbrace{g_i^\vee, \dots, g_i^\vee}_\text{$N$ times}
\Big) \qquad \text{for}\ i=1,2,3\,,
\\ \label{resi2}
L_0&\sim \mathrm{diag} \Big(
-g_0^\vee - 2(N-1)g, \ 
\underbrace{-g_0^\vee+2g, \dots, -g_0^\vee+2g}_\text{$N{-}1$ times}, \ \underbrace{g_0^\vee, \dots, g_0^\vee}_\text{$N$ times}
\Big)\,.
\end{align}

\medskip

In \cite{Tak99}, the Lax pair $L, A$ was constructed by an {\it ad hoc} method, and only for the Hamiltonian flow corresponding to the quadratic Hamiltonian $h_2$. A more general conceptual method for calculating $L, A$ was suggested in \cite{Chalykh_2019}. It uses elliptic Dunkl operators \cite{BFV94, EFMV} and, apart from reproducing the above $L$, it allows to construct a Lax partner $A$ for each of the commuting Hamiltonian flows. This means that $L$ remains isospectral under each of the flows governed by $h_2, h_4, \dots, h_{2N}$, cf. \cite[Prop. 5.6]{Chalykh_2019}.  As a result, the quantities $f_i=\tr(L^i)$ Poisson-commute with each of $h_{2k}$, hence $f_i$ is a function of $h_2, \dots, h_{2N}$. Taking into account \eqref{tran}, we conclude that each of the functions $f_i=\tr(L^i)$ is a polynomial in $h_2, \dots, h_{2N}$ whose coefficients are elliptic functions of $\alpha$. Hence, the characteristic polynomial of $L$ can be written as 
\begin{equation}
\det (L-k\mathrm{I})=k^{2N}+a_1k^{2N-1}+\dots+a_{2N}\,,
\end{equation}
where $a_i$ are polynomials in $h_2, \dots, h_{2N}$, elliptic in the spectral parameter.

\subsection{Spectral curve}

This puts us in the familiar setting of complex completely integrable systems. Namely, the level sets of $N$ Poisson-commuting Hamiltonians $h_2, \dots, h_{2N}$ define a Lagrangian fibration $\pi\,:\,\C^{2N}\to \C^N$. In addition to that, we have a family of {\it spectral curves} 
\begin{equation}\label{sc}
f(k, \alpha):=\det(L(\alpha)-k\mathrm{I})=0\,.
\end{equation}
parametrized by the coordinates $h_2, \dots, h_{2N}$ on the base of the fibration $\pi$. Each spectral curve \eqref{sc} is a $2N$-sheeted branched covering of the base elliptic curve $\Gamma=\C/(\Z+\tau\Z)$, with $(k, \alpha)$ viewed as coordinates on the cotangent bundle $T^*\Gamma$. The curve \eqref{sc} comes with a meromorphic differential, obtained by restriction from the canonical $1$-form $kd\alpha$ on $T^*\Gamma$, and a line bundle $\mathcal L$ (eigen-bundle of $L$). 
 
So far this looks parallel to the case of the usual Calogero--Moser system \cite{Krichever80}. Motivated by \cite{Hitchin, GN95, N96}, one should think of the matrix-valued $1$-form $\Phi:=L(\alpha)d\alpha$ as a Higgs field of some kind, so let us sketch such an interpretation. First, instead of considering $\Phi$ over the elliptic curve $\Gamma$, it is more natural to take into account the symmetry \eqref{sym}. It implies that 
\begin{equation}\label{fsym}
f(-k, -\alpha)=f(k,\alpha)
\end{equation} 
and so the spectral curve can be viewed as a branched covering of the Riemann sphere $\G/\Z_2$, with the $\Z_2$ acting by $\alpha\mapsto -\alpha$. Indeed, if we multiply $f(k,\alpha)$ by $(\wp'(\alpha))^{2N}$, we get
\begin{equation}\label{wf}
\widetilde f:=(\wp'(\alpha))^{2N}f(k, \alpha)=\det(\wp'(\alpha) L(\alpha)-k\wp'(\alpha)\mathrm{I})=\det(\widetilde L- y\mathrm{I})\,,
\end{equation}
where $\widetilde L=\wp'(\alpha)L$ and $y= k\wp'(\alpha)$. A quick check confirms that $\widetilde L$ is regular at $\alpha=\omega_r$, $r=1,2,3$, and that $\widetilde L(-\alpha)=M\widetilde L(\alpha) M^{-1}$. Therefore, the expression \eqref{wf} is a polynomial in $y$, whose coefficients are even elliptic functions with the only singularity at $\alpha=0$. As a result, the spectral curve \eqref{sc} acquires polynomial form
\begin{equation}\label{sc1}
\til f(x, y)= 0\,, \quad\text{where}\ x=\wp(\alpha)\,,\ y= k\wp'(\alpha)\,.
\end{equation}  
Using $x=\wp(\alpha)$ as the coordinate on $\Gamma/\Z_2$, we also obtain $\Phi=Ld\alpha= (\wp'(\alpha))^{-1}Ldx$. The properties of $L$ tell us that such $\Phi$ should be viewed as a Higgs field on the Riemann sphere with four marked points, more precisely, on an orbicurve $\CP^1$ of type $(2,2,2,2)$. Recall \cite{NS95} that Hitchin systems on orbicurves can also be viewed as parabolic Hitchin systems, with (conjugacy classes of) the residues of $\Phi$ at the marked points being associated with the values of the moment map, cf. \cite{N96, Donagi97}. Therefore, the formula \eqref{lm1} should be interpreted as a parametrization, by $p_i, q_i$, of the corresponding $2N$-dimensional symplectic leaf of a parabolic $\SL(2N, \C)$ Hitchin system on the Riemann sphere with four marked points $e_i=\wp(\omega_i)$, $i=0, 1, 2, 3$. This provides an interpretation of the Inozemtsev system as a Hitchin system. Note that this is different from the approach of \cite{HM01}.  Note also that the pattern \eqref{resi1}--\eqref{resi2} of the residues of $\Phi$ at the marked points is in good agreement with the SCFT picture (see Sec. \ref{secSWcrv} below). Also, as is explained below in Sec. \ref{sing}, the genus of the spectral curve equals $N$, which is as expected from both the Hitchin-system and the M5-brane perspectives.  

Let us also recall that starting from a moduli space $\mathcal M$ of Higgs bundles, the nonabelian Hodge correspondence and Riemann--Hilbert map associate to $\mathcal M$ two other moduli spaces, of local systems and of monodromy data (known as de Rahm and Betti models, see \cite{Boalch17} for a nice overview). For our case, these two other incarnations can be found in \cite{EGO05, Boalch11}, see also \cite{LO00, Tak01, Zotov, Kawakami15, BCR18} for further links between the Inozemtsev system and isomonodromic deformations.

\subsection{Spectral curves for $N=1$ and $N=2$}
\label{spec12}

Here we present explicit equations for the spectral curves \eqref{sc} in the cases of $N=1$ and $N=2$. We write equations in terms of the variables $k, \alpha$. They will be matched to M5 brane curves in Section \ref{section: BCtoM5_1}.

\subsubsection{$N=1$ curve}

For $N=1$, the Lax matrix is (cf. \cite{Zotov})
\begin{equation}
L= \begin{bmatrix}
p & v_\alpha(q)  \\
v_\alpha(-q) & -p 
\end{bmatrix}\,.    
\end{equation}
Using \ref{addv}, we find: 
\begin{equation}
\det L=-p^2-v_\alpha(q)v_\alpha(-q)=-p^2+u(q)-u^\vee(\alpha)\,, 
\end{equation}
where $u^\vee(\alpha)$ is the dual version of \eqref{u}, defined above in \eqref{u-vee}.
Hence, the spectral curve \eqref{sc} takes the form
\begin{equation}\label{BC1curve}
f(k,z) = k^2 - h_2 -u^\vee(\alpha)=0\,,
\end{equation}
with $h_2=p^2-u(q)$ viewed as a complex parameter. Multiplying this by $(\wp'(\alpha))^2$ and using $x=\wp(\alpha)$, $y=k\wp'(\alpha)$ we obtain $y^2=\wp'^2(\alpha)\, \left( h_2+u^\vee(\alpha) \right)$.
Using \eqref{wpidents} it is easy to see that the right-hand side is a quartic polynomial in $x=\wp(\alpha)$ (it reduces to a cubic if $g_0^\vee=0$). For generic $h_2$, the curves are smooth of genus $1$.

The Lagrangian fibration $\pi\,:\,\C^2\to\C$ is by the level sets $p^2-u(q)=h_2$. Singular fibers correspond to the stationary values of the Hamiltonian, i.e. to the equilibria $(p,q)=(0, q_0)$ with $u'(q_0)=0$.    
Then we can find that for a number of $l \geq 1$ generic couplings $g_i$, the number of stationary values of $h_2$ is $l+2$, in agreement with the Seiberg-Witten geometry \cite{SWii}. Indeed, the function $u'(q)= \sum_{i=0}^{3} g_i^2 \wp'(q+\omega_i)$ is odd elliptic of order $3l$, therefore it has $3l$ zeros; the genericity assumption ensures that the multiplicity of each zero is always one. Then $4-l$ zeros are given by the half-periods, for which the values of $h_2$ are distinct. Furthermore, the other $4l-4$ zeros come in pairs $(q, -q)$ so give the same stationary value of $h_2$. Thus, the number of singular fibers (or stationary values of $h_2$) is $(4-l) + (4l-4)/2 = l+2 $, as claimed.

\subsubsection{$N=2$ curve}

For $N = 2$, the Lax matrix is 
\begin{eqnarray}\label{n2mat}
L & = & \begin{bmatrix}
p_1 & g \sigma_\alpha(q_{12} ) & v_\alpha(q_1) &  g \sigma_\alpha(q^{+}_{12} ) \\
g \sigma_\alpha(-q_{12} ) & p_2  & g \sigma_\alpha(q^{+}_{12} )&  v_\alpha(q_2)\\
v_\alpha(-q_1) & g \sigma_\alpha(-q^{+}_{12} ) & -p_1 & g \sigma_\alpha(-q_{12} )\\
g \sigma_\alpha(-q^{+}_{12} )& v_\alpha(-q_2) & g \sigma_\alpha(q_{12} ) & -p_2
\end{bmatrix} \\
& = & P\begin{bmatrix}
p_1 &  v_\alpha(q_1) & g \sigma_\alpha(q_{12} ) & g \sigma_\alpha(q^{+}_{12} ) \\
v_\alpha(-q_1) &  -p_1 & g \sigma_\alpha(-q^{+}_{12} ) & g \sigma_\alpha(-q_{12} )\\
g \sigma_\alpha(-q_{12} ) &  g \sigma_\alpha(q^{+}_{12} )& p_2  & v_\alpha(q_2)\\
g \sigma_\alpha(-q^{+}_{12} )& g \sigma_\alpha(q_{12} ) &  v_\alpha(-q_2) & -p_2
\end{bmatrix}P^{-1}\,, \nn
\end{eqnarray}
where  
\begin{align}\label{Pperm}
P= \begin{bmatrix} 
1&0&0&0\\
0&0&1&0\\
0&1&0&0\\
0&0&0&1
\end{bmatrix} .
\end{align}

The $N=2$ case is the first case with non-zero ``antisymmetric mass" (related to the coupling $g$). If we let $g=0$, we find that the Lax matrix reduces to two $2 \times 2$ blocks, each having the form of a $N=1$ Lax matrix. Similarly, the general $2N\times 2N$ Lax matrix in the $g\to 0$ limit reduces to $N$ diagonal $2 \times 2$ blocks. Subsequently, in this limit the spectral curve is reducible, as it becomes a product of $N$ copies of the $N=1$ curve. 

The $N=2$ spectral curve is given by 
\begin{align}\label{BC2crv}
0=& (k^2 -u^\lor)^2 - h_2 (k^2 -u^\lor) + h_4  \\
& \quad 
- 4g^2 \Bigl( \wp(\alpha) (k^2 - u^\lor) 
+ g^\lor_0 \wp'(\alpha) k  
+ 2 (g^\lor_0)^2 \wp(\alpha)^2 
+ \wp(\alpha) \sum_{r=1}^3 (g^\lor_r)^2 \wp(\omega_r) \Bigr)\,,
\nn
\end{align}
where $u^\vee:=u^\vee(\alpha)$ and $h_2, h_4$ represent the values of two commuting Hamiltonians.   

The derivation of \eqref{BC2crv} is outlined in appendix \ref{appB}.

\subsection{Behaviour near marked points}\label{sing}

In order to make a connection with the analysis of the Seiberg--Witten curve in Sec. \ref{secSWcrv}, it will be useful to look more closely at the singularities of the Lax matrix \eqref{lm1}. This will also allow us to confirm that the genus of the spectral curves equals $N$, as expected. 

Expanding $L$ at half-periods gives
\begin{equation}
L=\sum_{j\ge -1}L_i^{(j)}(\alpha-\omega_i)^j\,,\quad i=0, 1, 2, 3\,,
\end{equation}
for some $L_i^{(j)}\in\mathrm{Mat}(2N, \C)$ independent of $\alpha$, with $L_i^{(-1)}$ being the residue matrices \eqref{res1}--\eqref{res2}. The property \eqref{sym} implies that 
\begin{equation}\label{mcom}
ML_i^{(j)} + (-1)^{j}L_i^{(j)}M=0\,,\qquad M = \begin{bmatrix}
0 & \mathrm{I}_N  \\ \mathrm{I}_N & 0 \end{bmatrix}\,.
\end{equation}
Now consider the $2N$ sheets of the spectral curve $\det (L-k\mathrm{I})=0$ near one of the half-period $\alpha=\omega_{1,2,3}$. From \eqref{resi1}, we know that locally we can label these sheets so the roots $k_1, \dots, k_{2N}$ near $\alpha=\omega_{i}$ behave as follows:
\begin{align}\label{exp1}
(k_1, \dots, k_{2N}) &\sim \frac1{\a-\om_i} \Big(
\underbrace{-g_i^\vee, \dots, -g_i^\vee}_\text{$N$ times}, \ 
\underbrace{g_i^\vee, \dots, g_i^\vee}_\text{$N$ times}
\Big)+\text{regular terms}\,.
\end{align}
Series expansions for each $k_r(\alpha)$ can be worked out recursively, as a perturbation series, together with the eigenvectors $v_r(\alpha)$ such that
\begin{equation}
L(\alpha)v_r(\alpha)=k_r(\alpha)v_r(\alpha)\,,\qquad v_r(\alpha)=\sum_{j\ge 0}v_r^{(j)}(\alpha-\omega_i)^j\,,
\end{equation}   
for a chosen ``initial" eigenbasis $v_r^{(0)}$ of the residue matrix $L_i^{(-1)}$. Since the residue matrix $L_i^{(-1)}$ commutes with $M$ for all $i=0,1,2,3$ (for $i\ne 0$ it is simply proportional to $M$), the chosen eigenvectors are also eigenvectors of $M$, and so half of them satisfy $Mv_r^{(0)}=v_r^{(0)}$, with $Mv_r^{(0)}=-v_r^{(0)}$ for the other half. The additional symmetry \eqref{mcom} of the Lax matrix imposes extra constraints, which result in the following:

\noindent{\bf 1.} Near $\alpha=\omega_i$, each eigenvalue $k_r(\alpha)$ is odd, i.e. it changes sign under $\alpha\mapsto 2\omega_i-\alpha$.

\noindent{\bf 2.} The terms of the series for the eigenvector $v_r(\alpha)$ satisfy $Mv_r^{(j)}=\pm (-1)^jv_r^{(j)}$, with the sign $\pm$ determined by the initial eigenvector $v_r^{(0)}$.

An important corollary of the first property is that the regular terms in \eqref{exp1} are in fact of order $O(\alpha-\omega_i)$.
Then by squaring the spectral variable $k$ and shifting it appropriately, all the poles can be cancelled.  In particular, 
\begin{align}\label{zreg}
z &\sim \frac1{(\a-\om_i)^2} \Big(  
\underbrace{0, \dots ,0}_{2N\ \text{times}} 
\Big)+\text{regular terms}\quad(i=1,2,3)\,,
\end{align}
where we have defined
\begin{align}\label{u-vee}
z &:=\frac14 \left( k^{2} -u^\lor + \text{constant}\right) ,&
u^\vee &= \sum_{i=0}^3 (g_i^\vee)^2 \wp(\a + \om_i)\,.
\end{align}
The factor of $1/4$ and the constant in \eqref{u-vee} are for later convenience.

The same analysis for $\alpha\sim 0$ gives that
\begin{align}
(k_1, \dots, k_{2N}) &\sim \frac1{\a}  \Big(
-g_0^\vee - 2(N-1)g, \ 
\underbrace{2g-g_0^\vee, \dots, 2g-g_0^\vee}_\text{$N{-}1$ times}, \ \underbrace{g_0^\vee, \dots, g_0^\vee}_\text{$N$ times}
\Big)+{O}(\alpha)\,,
\end{align} 
 and so by squaring and shifting it appropriately all but one of the $2N$ poles there can be cancelled.  In particular, 
\begin{align}\label{ztilpole}
\til z := z + \frac gx \left(y + \frac12 g^\lor_0 x^2 \right) 
&\sim \frac1{\a^2} \Big(
Ng (g^\lor_0 + (N-1) g) , \ 
\underbrace{0,\dots ,0}_{2N{-}1\ \text{times}} 
\Big)+\text{regular terms}\,,
\end{align}
where we have defined
\begin{align}\label{xydef}
x := \wp(\a) \sim \frac1{\a^2}\,,\qquad 
y :=\frac14 k\wp'(\a) .
\end{align}
\eqref{ztilpole} indicates that the coefficients of the spectral curve written in the $(x,y,\til z)$ variables (as an $N$-fold cover of the sphere parametrized by $x$) can only have simple poles at $x=\infty$, while \eqref{zreg} indicates that if they are written in the $(x,y,z)$ variables they will be regular away from $x=\infty$.  In fact this observation will play an important role in finding the change of variables needed to match the spectral curve to the Seiberg-Witten curve, discussed in section \ref{secSWcrv}.

We can now calculate the genus of the spectral curve \eqref{sc1}. We follow the same method as in \cite{Krichever80}. First, consider the curve $\Gamma_N$ \eqref{sc} and denote its genus  by $g$. Then $2g-2=\nu$,where $\nu$ is the number of the branch points of $\Gamma_N$ viewed as a covering of the elliptic curve $\Gamma$. This is the number of zeros of $\partial f/{\partial k}$ on $\Gamma_N$; it also equals the number of poles of $\partial f/{\partial k}$. The poles occur precisely at $2N$ points of $\Gamma_N$ above each of the half-periods $\alpha=\omega_i$. Locally, we can factorize $f(k, \alpha)$ into a product of factors $k-k_r(\alpha)$. For example, near $\alpha=\omega_{1,2,3}$ we have
\begin{equation}
f(k, \alpha)=\prod_{r=1}^N \left(k+\frac{g_i^\vee}{\alpha-\omega_i} + b_r(\a)\right)\prod_{r=N+1}^{2N}\left(k-\frac{g_i^\vee}{\alpha-\omega_i} + b_r(\a)\right)\,,
\end{equation}  
where the $b_r(\a)$ are of order ${O}(\alpha-\omega_i)$.  By differentiating this equation with respect to $k$, we find that $\partial f/\partial k$ has a simple pole on each of the $2N$ sheets above $\omega_i$. A similar analysis near $\alpha=0$ shows that $\partial f/\partial k$ has there a pole of order $2N-1$ on one sheet,  poles of order $3$ on $N-1$ sheets, and simple poles on the remaining $N$ sheets. This gives $2g-2=3\times 2N+(2N-1)+3\times (N-1) +N =12N-4$, so $g=6N-1$.  

The curve $\Gamma_N'$ \eqref{sc1} is obtained from $\Gamma_N$ by taking a quotient by the involution $(k, \alpha)\mapsto (-k, -\alpha)$. Thus, $\Gamma_N$ can be viewed as a $2$-sheeted covering of $\Gamma_N'$, branched at the fixed points of the involution. These are precisely the points above the half-periods, so there are $8N$ of them. Denoting by $g'$ the genus of $\Gamma_N'$, we get $12N-4=2g-2=2(2g'-2)+8N$, from which $g'=N$, as claimed.

\subsection{Modular property}\label{section:modular}

The Lax matrix and the spectral curve exhibit a modular behaviour under $\SL(2, \Z)$-action. To state the result, recall that the Lax matrix $L$ depends on the modular parameter $\tau$, the spectral parameter $\alpha$, $2n$ variables $p_i, q_i$, and the coupling constants $g$ and $g_{0,1,2,3}$. Take $\g =\left(\bsm a & b\\ c & d \esm\right) \in \SL(2,\Z)$ and define $L'$ to be the Lax matrix with the variables changed to $\tau'$, $\alpha'$, etc., in the following way:
\begin{align}
\t' & = \frac{a\t +b}{c\t +d}\,,\quad \alpha'  = (c\t+d)^{-1}\alpha\,,\\
p_i' &= (c\t+d) p_i\,,\quad q_i' = (c\t+d)^{-1} q_i\,,\\
g' & =g\,,\quad g_0'=g_0\,,\quad g'_r = g_{\pi_{\g}(r)}\quad \text{for}\ r=1,2,3\,.
\end{align}
Here in the last formula we denote by $\pi_\gamma$ the permutation of $\{1,2,3\}$ determined by the group homomorphism \eqref{hom}.
With this notation, we have:
\begin{align}\label{mpr}
L' = (c\t+d) Q L Q^{-1}\,,
\end{align}
where 
$Q= \begin{bmatrix} R & 0 \\ 0 & R^{-1} \end{bmatrix}$ and $R = \mathrm{diag}\left(\exp(-\frac{2\pi ic}{c\t+d} \alpha q_1), \dots, \exp(-\frac{2\pi ic}{c\t+d} \alpha q_N)\right)$.

The formula \eqref{mpr} is obtained in a straightforward way from the modular properties of the functions $\sigma_\alpha(x)$ and $v_{\alpha}(x)$ given in the Appendix. If we introduce $k'=(c\t+d) k$, then we also have 
\begin{equation}
\det(L'-k'\mathrm{I})=(c\tau +d)^{2N}\det(L-k\mathrm{I})\,.
\end{equation}
The physical interpretation of these properties on the QFT side is the $\SL(2,\Z)$ S-duality mixed with the Spin(8) triality (see Sec. \ref{section:3.1} below).

\section{$\USp(2N)$ $N_f=4$ superconformal field theory}\label{section:D4}

We consider the family of 4d $\cN=2$ superconformal field theories consisting of $\USp(2N)$ gauge theories with $N_f = 4$ hypermultiplets in the fundamental representation and (for $N \geq 2$) $N_a=1$ hypermultiplets in the traceless antisymmetric two-index tensor representation. 

\subsection{Field theory properties}\label{section:3.1}

We list some long-established properties of these theories.
\begin{itemize}
	\item They are a family of interacting 4d $\cN=2$ SCFTs labelled by a positive integer $N$, which we call the rank of the $N_f=4$ theory.  As SCFTs, they are invariant under the 4d $\cN=2$ superconformal group $\SU(2,2|2)$.
	\item The $N_f=4$ SCFTs have an exact $\SL(2,\Z)$ S-duality.  This means that each theory has a one-complex-dimensional conformal manifold given by the upper half complex plane modulo $\SL(2,\Z)$ M\"obius transformations.  Though the center of $\SL(2, \Z)$ acts trivially on the conformal manifold, it acts non-trivially as charge conjugation in the field theory.  Around a special point on the conformal manifold the theory admits a weakly-coupled Lagrangian description in terms of $\USp(2N)$ gauge theory with 4 fundamental and 1 antisymmetric hypermultiplets.  The weak coupling limit of the complex gauge coupling constant $\t$ parameterizing the conformal manifold is $\mathrm{Im}(\t) \to \infty$.
	\item The internal global ``flavor" symmetry is $\Spin(8)$ for $N=1$ and $\Spin(8) \times \SU(2)$ for $N \geq 2$, under which the four fundamental hypermultiplets (the same as eight fundamental half-hypermultiplets) transform in the $(8_v,1)$ representation, and the antisymmetric hypermultiplet in the $(1, 2)$ representation.  Correspondingly, there is a space of $\cN=2$-preserving mass deformations given by the complexified weight space of $\Spin(8) \times \SU(2)$.  Introduce mass (or deformation) parameters $(m_1,\ldots,m_4)$ for $\Spin(8)$ and $M$ for $\SU(2)$ as linear coordinates on this parameter space such that $m_i$ is the complex mass of the $i$-th fundamental hypermultiplet, and $M$ the mass of the antisymmetric hypermultiplet.\footnote{We use an unconventional normalization for the mass such that our masses $m$ are related to the canonically normalized masses $\til m$ by $\til m = \sqrt2 \, m$.}  The principal congruence subgroup $\G(2)\subset \SL(2,\Z)$ of the S-duality group acts trivially on the $\Spin(8)$ masses, while the quotient $\SL(2, \Z)/\G(2) \simeq S_3$ transforms the mass parameters by the $\Spin(8)$ ``triality" outer automorphism \cite{SWii, LY97}.  The antisymmetric mass is invariant under S-duality transformations.
	\item The operator content of an $N_f=4$ theory can be organized in terms of the unitary representations of its global symmetry $\SU(2,2|2) \times \Spin(8) \times \SU(2)$. In particular, with respect to $\SU(2,2|2)$ there are various sectors of supersymmetry-protected BPS operators, for instance, Coulomb branch operators and Higgs branch operators. The condensate of the scalar components in the $\cN=2$ multiplets of BPS operators parameterize moduli spaces of $\cN=2$ invariant vacuum states. 
	\item The moduli space of vacua consists of various branches each of which is locally a metric product of a Coulomb factor and a Higgs factor, with complex dimension $n_C$ and quaternionic dimension $n_H$, respectively.  Conventionally, the branch with maximal $n_C$ is called the Coulomb branch and the branch with maximal $n_H$ the Higgs branch. The rank $N$ $N_f=4$ theory has a Coulomb branch with $(n_C, n_H) = (N, N-1)$ and a Higgs branch with $(n_C, n_H) = (0, 6 N-1)$.  The $N-1$ quaternionic dimensional Higgs factor of the Coulomb branch comes from the components of the antisymmetric hypermultiplet carrying zero weight with respect to the $\USp(2N)$ gauge algebra.
	\item The vector multiplet of the Lagrangian theory contains a scalar field $\Phi$ in the adjoint representation.  The Coulomb branch coordinate ring is freely generated by $u_i := \tr(\wedge^{2i}\Phi)$ with $i = 1, 2, \ldots, N$, corresponding to the primitive Casimir elements of $\USp(2N)$.  The Coulomb branch coordinate ring is graded by the scaling dimension, so the weight of $u_i$ is $2i$.   Since the Coulomb branch chiral operators are BPS operators, this description of the Coulomb branch chiral ring is true at all points of the conformal manifold, not just at the weak coupling point.
\end{itemize}

We are interested in the geometry of the Coulomb branch.  The low energy effective $\U(1)^N$ gauge theory on the Coulomb branch is encoded in the special K\"ahler geometry \cite{Freed97} of the Coulomb branch.  The $N-1$ massless neutral hypermultiplets on the Coulomb branch decouple in the low energy limit, so will be ignored.  

On general grounds \cite{DW95} a Coulomb branch special K\"ahler geometry is equivalent to a classical complex completely integrable Hamiltonian system.  In particular, the Coulomb branch is the $N$-complex-dimensional manifold of the action variables of the integrable system.  The matrix of low energy $\U(1)^N$ complex gauge couplings gives the period matrix of a complex torus of dimension $N$, so the Coulomb branch parameterizes a family of complex tori, giving the angle variables of the integrable system. The complex tori are also endowed with principle polarization coming from the Dirac pairing on the $\U(1)^N$ electric-magnetic charge lattice, and hence are abelian varieties. The total space of this family of abelian varieties is a complex symplectic variety, the complex phase space of the integrable system, with holomorphic symplectic form $\om$.

The next subsection describes the total space geometry by way of a holomorphic family $\Sigma$ of genus-$N$ Riemann surfaces over the Coulomb branch, along with a meromorphic one-form $\l$ on the fibers whose poles have constant residues.  $(\Sigma,\l)$ are called the Seiberg-Witten curve and one-form in the physics literature.  The abelian variety fibers of the integrable system are the Jacobian tori of the Riemann surfaces, and the symplectic form is $\om= d\l$.   Thus we will match the field theory Coulomb branch geometry to the Inozemtsev system by matching the Seiberg-Witten curve and one-form to the spectral curve and canonical one-form of the integrable system.

\subsection{Seiberg-Witten curve}\label{secSWcrv}

The $\USp(2N)$ $N_f=4$ SCFTs can be constructed as the low energy effective theory of type IIA superstrings in the presence D4, NS5, D6, and O6$^-$ branes generalizing the construction of \cite{Witten97}. The M-theory lift of the D6 and O6$^-$ IIA brane configuration \cite{GK98} is a specific choice of complex structure of a $(T^2 \times \C)/\Z_2$ hyperk\"ahler orbifold background.  The M-theory lift of the D4 and NS5 branes is a single M5 brane intersecting the background except over points of $T^2 $ corresponding to NS5 branes.  This intersection is the Seiberg-Witten curve, and the restriction of the holomorphic hyperkahler form to the curve is the Seiberg-Witten one-form.  This is the spectral curve of a Hitchin system on the orbifolded torus with punctures \cite{GK98}.  

The deformations of this orbifold background and M5 brane curve corresponding to turning on the $\Spin(8)$ fundamental masses and the $\SU(2)$ antisymmetric mass was worked out in \cite{AMP02}.  The connection to a Hitchin system is no longer apparent in this description.  We will describe this solution for the $\USp(2N)$ $N_f=4$ Coulomb branch in more detail shortly in preparation for showing its equivalence to the spectral curve of the Inozemtsev system.  But first, we make a few comments on two other string constructions of the $\USp(2N)$ $N_f=4$ theories.

These theories naturally arise as the world volume theories on a stack of $N$ parallel D3 branes probing an F-theory singularity of $(I^\ast_0, D_4)$ type --- i.e., an $\mathrm{O7}^-$ plane coinciding with four $\mathrm{D7}$ branes \cite{Sen96, BDS96, ASYT96, DLS96}.  But it is not known how to turn on the antisymmetric mass $M$ deformation in the F-theory construction. 

These theories also admit a class-S construction via a 6d $(2,0)$ $A_{2N-1}$ SCFT compactified on a sphere $C$ with four punctures all of type $[N,N]$.  This construction only makes manifest an $\SU(2)^4$ subgroup of the $\Spin(8)$ flavor group, and does not make the antisymmetric $\SU(2)$ flavor factor or its associated mass deformation apparent \cite{NX09}.  $C$ is identified with $T^2/\Z_2$ with the four punctures corresponding to the four $\Z_2$ orbifold fixed points.  The antisymmetric hypermultiplet appears upon taking an appropriate zero-area limit of $C$ \cite{GMT11}, and \cite{GR12} showed that by modifying the type of one puncture to be $[N, N-1, 1]$, the theory manifests the antisymmetric $\SU(2)$ flavor symmetry.  The class-S construction realizes the integrable system underlying the Coulomb branch geometry as a Hitchin system \cite{GMN09}.

The matching to the M5 brane curve, presented below, gives strong evidence that the Hitchin system associated with the above class-S construction can be identified with the Inozemtsev system.

In the rest of this section we review the M5 brane construction \cite{AMP02} of the SW curve for the $\USp(2N)$ $N_f=4$ theory.  The main ingredients in this construction are:
\begin{itemize}
\item The $\USp(2N)$ theory with the $\Spin(8)$ mass deformation is realized by embedding one complex dimension of the M5 brane world volume in a complex surface, $Q_0$.  $Q_0$ carries a hyperk\"ahler structure --- from which the SW 1-form is derived --- and is a deformation of a $(T^2\times\C)/\Z_2$ orbifold.  This surface can be thought of (we will be more precise below) as fibered over $T^2/\Z_2$.
\item The intersection with the M5 brane then gives a curve which projects to an $N$-fold cover of $T^2/\Z_2$ minus one of the orbifold points.   At the missing orbifold point the M5 brane is not transverse to $Q_0$; we will call this point the ``pole" of the M5 brane.
\item The $\SU(2)$ mass deformation, $M$, is realized by further deforming the background surface to $Q_M$.  Following the discussion of the analogous deformation of the elliptic model in \cite{Witten97}, describe $Q_M$ by two charts to $Q_0$, one including the fibers above a neighborhood of a chosen point $p \in T^2/\Z_2$, and the other encompassing the rest of the surface.  The two coordinate patches are isomorphic to the corresponding patches of $Q_0$, and the $M$ deformation is realized by requiring that the transition map is a shift of the fiber coordinate which has a pole with residue proportional to $M$ at $p$.  We call this transition map the ``$M$ shift".  Changing $p$ and the form of the transition map but keeping $M$ fixed does not change the complex structure of $Q_M$.
\item The M5 brane curve for the mass-deformed $\USp(2N)$ $N_f=4$ SCFT is then locally a degree-$N$ polynomial in the fiber coordinate on $Q_M$ whose coefficients have at most a simple pole over a chosen orbifold point of $T^2/\Z_2$.
\end{itemize}

The form of the SW curve for the $\USp(2N)$ $N_f=4$ (and many other closely related) SCFTs found in \cite{AMP02} followed this procedure with the $M$ shift at a point $p$ not equal to one of the orbifold points of $T^2/\Z_2$.  Both the form of the spectral curve of the Inozemtsev system as well as the above-mentioned S-class construction (where one of the four punctures is modified to capture the $M$ deformation) suggest that they will most easily match the form of the SW curve if the point $p$ of the $M$ shift should be taken to coincide with one of the orbifold points.  This involves a slight modification of the construction of \cite{AMP02} which we now explain.

\subsubsection{Background surface}

We start with the orbifold $(T^2 \times \C)/\Z_2$.  Think of $T^2\times \C$ as an affine bundle over $T^2$ and let $v \in \C$ be the fiber coordinate.  Write the complex torus $T^2$ as a curve $\eta^2 = \prod_{i=1}^4 (x-e_i w)$ in weighted projective space, $[w:x:\eta]\in\P^2_ {(1,1,2)}$.  Note that $\SL(2,\C)$ transformations of $(w,x)$ do not change the complex structure of $T^2$, but change the $e_i$ by M\"obius transformations.  The $\Z_2$ identification on $\C \times T^2$ is $(v,w,x,\eta) \simeq (-v,w,x,-\eta)$.  Using the invariant coordinates on the orbifold, $y = v\eta$, $z = v^2$ ($w$ and $x$ unchanged), the orbifolded background space is given by the surface $y^2 = z \prod_{i=1}^4 (x-e_i w)$. 

The $(T^2 \times \C)/\Z_2$ orbifold has a four-parameter deformation into a complex surface $Q_0$ with the same asymptotic structure. The mass-deformed orbifold surface $Q_0$ and SW 1-form are \cite{AMP02}
\begin{align}\label{crv1.5} 
\l &= \frac{y(wdx-xdw)}{P}  ,&
P &:= \prod_i (x-e_iw) ,
\nn\\
y^2 &= z P + Q,&
Q &:= \sum_j \mu_j^2 w  \prod_{k\neq j} [(x-e_kw)(e_j-e_k)] ,
\end{align} 
where $i,j,k\in\{0,1,2,3\}$.  Note that we still have $[w:x:y] \in \P^2_{(1,1,2)}$.
The deformation parameters, $\m_i$, turn out to be related to the fundamental masses by \cite{AMP02}
\begin{equation}
\m_0 = \tfrac12 (m_1 + m_2), 
\quad \m_1 = \tfrac12 (m_1 - m_2), 
\quad \m_2 = \tfrac12 (m_3 + m_4), 
\quad \m_3 = \tfrac12 (m_3 - m_4) .
\end{equation}

The topology of $Q_0$ can be pictured by noting that the $z =$ constant ``sections'' are tori, and the $x= \x w$ ($\x=$ constant) ``fibers'' are generically 2-sheeted covers of the $z $-plane branched over the point $z = -Q/P$.  But when $x=e_i w$ the fiber becomes two disconnected copies of the $z $-plane,
$S^\pm_j := 
\bigl\{\, x = e_j w, \ y = \pm \m_j w^2 {\textstyle \prod_{k\neq j}} (e_j-e_k), \ \forall z  \, \bigr\}$.
The existence of these ``double fibers" over the Weierstrass points in the deformed orbifold will play a central role in what follows.  From the point of view of the IIA string theory D4/NS5/O6$^-$ brane construction, the generic $x=\x w$ fibers correspond to possible loci of (the M theory lift of) an NS5 brane, while the $S^\pm_j$ curves correspond the possible loci of ``half'' NS5 branes ``stuck'' at an O6$^-$ orientifold plane.

To get closer to the form of the integrable system spectral curve, we will specialize \eqref{crv1.5} to Weierstrass form where the Weierstrass points are placed at $e_0=\infty$ and $\sum_{j=1}^3 e_j =0$.  Then the $Q_0$ surface and 1-form become
\begin{align}\label{surfwp}
\l &= \frac{y (wdx-xdw)}{w\til P} , &
\til P &:= \prod_i (x-e_i w) 
= x^3 + s_2 w^2 x - s_3 w^3, \nn\\
y^2 &= (z w + \m_0^2 x) \til P + w^2 \til Q, &
\til Q & := \sum_j \m_j^2 \e_j \prod_{k\neq j} (x - e_k w) .
\end{align}
\red{}
where now indices only take the three values $i,j,k\in\{1,2,3\}$, and we have defined the useful combinations
\begin{align}\label{sepdef}
s_2 & := \sum_{j<k} e_j e_k, &
s_3 & := \prod_j e_j, &
\e_j &:= \prod_{k\neq j} (e_j-e_k) .
\end{align}
Note that the equations for the disjoint fibers over the Weierstrass points become
\begin{align}\label{wpysols}
S^\pm_\infty &:= \{ w=0, y = \pm \mu_0 x^2, \forall z  \},&
&\text{and}&
S^\pm_j &:= \{ x{=}e_j w,\, y{=}\pm \m_j \e_j w^2,\, \forall z  \}.
\end{align}

Now we discuss the $M$ deformation with the shift put at a branch point.  To motivate the construction, we first review, following \cite{Witten97}, the corresponding deformation of the unorbifolded $T^2\times \C$ background, $\eta^2 = P$.  Put the $M$ shift at the Weierstrass point $w=0$ (which is $x=\infty$ in the $w=1$ patch) by defining the transition map, 
\begin{align}\label{vshift}
\til v = v + M \frac{\eta}{w x} ,
\end{align}
where $\til v$ is the fiber coordinate of a chart over a neighborhood of the $w=0$ point of the $T^2$.  This transition map has a pole with residue $M$ over $w=0$, so describes a one-parameter complex deformation of $T^2\times \C$ with parameter $M$.  This is because the deformations of the affine bundle $T^2\times \C$ are classified by $H^1(T^2, \cO_{T^2})$ which is 1-dimensional, so there is just a single deformation parameter, and furthermore this cohomology group vanishes if a point is deleted from $T^2$.

In our case $Q_0$ is not an affine bundle, but is a deformation of a $\Z_2$ orbifold of the this affine bundle.  This leads to the expectation (for which we do not have a rigorous justification) that there is still only a single complex deformation preserving the asymptotic structure.  We can find a description of this deformation simply by orbifolding the $M$ shift given in \eqref{vshift}, or more generally, by defining the transition map to be any shift of the ``fiber'' ($z $) coordinate with a pole over the Weierstrass point $w=0$ with residue proportional to $M$.

The $\Z_2$ orbifold action identifies $\til v \leftrightarrow -\til v$, so we define invariant coordinates $\zt = {\til v}^2$, $\yt = \til v \eta$.  Then \eqref{vshift} gives the transition map
\begin{align}\label{vcov2}
\yt &= y + M \frac{\til P}{x} ,&
\zt &= z  + 2M \frac y{wx} + M^2 \frac{\til P}{w x^2} ,
\end{align}
in a neighborhood of the $w=0$ fiber of $(\C\times T^2)/\Z_2$.  Thus $y$ is shifted by a term regular at $w=0$ (in the $x=1$ patch), while $z $ is shifted by a double pole at $w=0$ plus single pole and regular terms.   (Recall that in local coordinates around $w=0$ --- i.e., $\sqrt w$ in the $x=1$ patch --- $y$ has a simple zero and $w^{-1}$ a double pole.)

So far this has all been in the undeformed orbifold.  To go to the $Q_0$ surface where the orbifold is deformed by turning on the $\m_i$ masses, it was argued in \cite{AMP02} that \eqref{vcov2} does not change, since one simply shifts $z  \to z  + \frac QP$ and the same for $\zt$.   In Weierstrass form this applies without change; just rewrite $\frac QP = \m_0^2 \frac xw + \frac{\til Q}{w\til P}$.

But \eqref{vcov2} has a qualitatively different pole structure at $w=0$ in $Q_0$ than in the undeformed orbifold. In the undeformed orbifold $y \sim \sqrt w$ was  the local coordinate vanishing at $w=0$, but in the deformed orbifold $w=0$ is no longer a branch point for $y$; instead $y$ has two solutions, giving two disjoint curves over $w=0$, denoted by $S^\pm_\infty$ in \eqref{wpysols}.  In the neighborhood of $S^\pm_\infty$ the transition map \eqref{vcov2} has a pair of distinct simple poles along $S^\pm_\infty$ rather than a single double pole.

Although the form of the $M$ shift given in \eqref{vcov2} is perfectly valid, the form of the resulting M5 brane curves do not match to those of the Inozemtsev system in an algebraically simple way.  Confident that there is only a single complex deformation $Q_0 \to Q_M$, we can modify \eqref{vcov2} to any other convenient transition map which has a simple pole in $\zt$ at $w=0$.  

The property \eqref{ztilpole} of the spectral curve indicates that $\zt$ should be chosen to have only a single pole at $w=0$ ($x=\infty$).  A simple transition map which does this is
\begin{align}\label{case2cov}
\yt &= y , &
\zt &= z  + 2M \frac{(y+\mu_0 x^2)}{wx} .
\end{align}
since \eqref{case2cov} behaves near $w=0$ as
\begin{align}\label{case2cov2}
\zt &= 
\begin{cases}
(1+ \tfrac{M}{\m_0}) z  + 2\m_0 M \frac{x}{w}  
&\text{at $S^+_\infty$} \\
(1 - \tfrac{M}{\m_0}) z 
& \text{at $S^-_\infty$},
\end{cases}
\end{align}
so has a simple pole only along the $S^+_\infty$ fiber over $w=0$, and is regular along the $S^-_\infty$ fiber.  

We will see below that this transition map gives an M5 brane curve which is easily matched to the Inozemtsev system spectral curve.  Indeed, comparing \eqref{u-vee}, \eqref{ztilpole} and \eqref{xydef} to \eqref{case2cov} already indicates how most of the variables and parameters of the integrable system will have to be matched to those of the SW curve.

\subsubsection{M5 brane curve}

We now have a choice of placing of a stuck NS5 brane at $w=0$ at either the $S^+_\infty$ or the $S^-_\infty$ fiber.  This choice gives two different forms of the curve upon turning on the $M$ deformation since it gives different regularity conditions in the shifted $\zt$ coordinates depending on whether the stuck brane coincides with the shift pole or not.  However, once again the property \eqref{ztilpole} of the spectral curve indicating that there should be only a single pole dictates that the stuck NS5 brane should be placed at the $S^+_\infty$ fiber to coincide with the position of the $M$ shift pole.

Before turning on the $M$ deformation, the M5 brane curve of \cite{AMP02} in the $Q_0$ background specialized to the case of the $\USp(2N)$ $N_f=4$ theory has the form $0 =  z ^N + \cA(w,x,y,z )$ where $\cA$ is a polynomial in $z $ of order $N-1$, homogeneous of weight 0 in $(w,x,y)$, and can have a simple pole along either the $S^+_\infty$ or the $S^-_\infty$ fiber over $w=0$.  This comes from the IIA brane construction where $N$ is the number of D4 branes (after orbifolding) corresponding to the rank of the gauge group and the pole at $w=0$ is a single stuck NS5 brane.  A linear basis of functions of $(w,x,y)$ homogeneous of weight 0 with at most a simple pole at $w=0$ is $\{ 1, x/w\}$.  Thus $\cA$ can be written more explicitly as
\begin{align}\label{Q0M5crv} 
0 &= z ^N + A_0(z ) + \frac{x}{w}A_1(z )
\end{align}
where $A_{0,1}$ are arbitrary polynomials of order $N-1$ in $z $.  Since the curve is allowed to have a pole only along either $S^+_\infty$ or $S^-_\infty$, but not along both, and since $\frac{x}{w}$ has a pole along both, we must, in fact, have that $A_1(z )\equiv 0$.  Thus, when $M=0$ the $\USp(2N)$ $N_f=4$ curve is generically $N$ disjoint sections of $Q_0$ corresponding to the $N$ roots of the polynomial $z ^N+A_0(z )$.  This reflect the well-known fact --- reviewed at the beginning of the next section --- that when $M=0$ the Coulomb branch of the theory is the  $N$-fold symmetric product of the rank-1 Coulomb branch.

We now turn on the antisymmetric mass deformation parameter $M$ by using the transition map \eqref{case2cov}.  Concretely, the curve for the shifted model is like the curve for the non-shifted model \eqref{Q0M5crv} except that we should now allow singularities only $S^+_\infty$ in a coordinate patch covering $w=0$ with coordinates $(w,x,y,\zt)$ related to $(w,x,y,z )$ by \eqref{case2cov}. 

Since we are only adding poles at $w=0$, and the only functions of weight zero in $(w,x,y)$ with poles only there are $(x/w)^\a$ and $(y/w^2)(x/w)^\a$ for non-negative $\a$, the general form of the curve in the $z $ patch will be
\begin{align}\label{F1}
0= F := z ^N + \sum_{a=0}^\infty 
\frac {w^2 A_a + y E_a}{w^2} \Bigl(\frac xw \Bigr)^a
\end{align}
where the $A_a$ and $E_a$ are arbitrary polynomials of order $N-1$ in $z $.

Though \eqref{F1} is a correct general form for the curve, the infinite sum of pole terms is intimidating.  It is not too hard to bound the number of pole terms that can contribute by using the condition that there is only at most a first-order pole at $w=0$ in the shifted $\zt$ variable.
Under the transition map \eqref{case2cov}, $\zt = z  + y P_1 + P_1$, where $P_a$ refers to a generic rational function of $w$ with poles of up to order $a$ at $w=0$ (work in the $x=1$ patch).  Using the fact that  $y^2 \sim z  P_0 + P_0$, one can recursively eliminate all higher powers of $y$ in $\zt^\ell \sim z ^\ell + \cdots$ to find that
\begin{align}\label{ztellpole}
\zt^\ell \lesssim z ^\ell + \sum_{a=1}^{\ell-1} z ^{\ell-a} (P_{2a} + y P_{2a-1}).
\end{align}
The $\lesssim$ sign means that we have pole orders bounded by the terms on the right.  
In the $\zt$ coordinate the curve is to have at most a simple pole at $w=0$, so will have the form $\zt^N + \sum_{\ell = 0}^{N-1} \zt^\ell P_1$.  Substituting \eqref{ztellpole} into this then shows that in the $z $ coordinate the highest-order poles are of the form
\begin{align}\label{Fpoleorder}
F & \ \lesssim\  z ^N + \sum_{\ell = 0}^{N-1} \sum_{a=0}^\ell z ^{\ell-a} (P_{2a+1} + y P'_{2a}) 
\ \sim\ z ^N + \sum_{\ell=1}^N z ^{N-\ell} (P_{2\ell-1} + y P'_{2\ell-2}),
\end{align}
where by $P'_a$ we mean the usual $a$th-order pole for $a\neq0$, but $P'_0\equiv0$.  Comparing to \eqref{F1} then implies that the curve is
\begin{align}\label{QMcur} 
0= z ^N 
+ \sum_{\ell=1}^N z ^{N-\ell} 
\Bigl( 
\sum_{a=0}^{2\ell-1}  A_{a\ell}\frac{x^a}{w^a}
+ \sum_{a=0}^{2\ell-2}  E_{a\ell}\frac{y x^a}{w^{a+2}}  
\Bigr) .
\end{align}
Note that \eqref{ztellpole} and thus \eqref{QMcur} does not give the optimal bound on the order of the poles appearing in the curve, but instead just gives a reasonable upper bound.  This is not a big deal since any ``extra" terms will be set to zero upon demanding only a simple pole appear in the $\zt$ patch.

The coefficients in \eqref{QMcur} are determined by demanding the correct pole behavior after shifting to the $\zt$ variable.  Concretely, make the inverse change of coordinates \eqref{case2cov} in the curve by  substituting $z  \to \zt - 2M (y+\m_0 x^2)/(wx)$ in \eqref{QMcur}.   
The 5 brane curve \eqref{QMcur} in the $x=1$ patch becomes in terms of the \eqref{case2cov} shifted variables
\begin{align}\label{case2Spcrv} 
0= \Bigl( \zt - 2M \frac{y+\mu_0}{w} \Bigr)^N 
+ \sum_{\ell=1}^N \Bigl( \zt - 2M \frac{y+\mu_0}{w} \Bigr)^{N-\ell} 
\Bigl( 
\sum_{a=0}^{2\ell-1}  \frac{A_{a\ell}}{w^a}
+ \sum_{a=0}^{2\ell-2}  \frac{y E_{a\ell}}{w^{a+2}}  
\Bigr) .
\end{align}
Expand this around $w=0$ keeping only pole terms $\zt^\ell w^{-a}$ and $\zt^\ell y w^{-a}$ for $a>0$. 
We do this by using iteratively that $y^2 = (\zt w - 2M (y+\mu_0) + \m_0^2) \til P + w^2 \til Q$, with $\til P = 1 + s_2 w^2 - s_3 w^3$, and $\til Q = \sum_j \m_j^2 \e_j \prod_{k\neq j} (1 - e_k w)$ to reduce all terms to either $\zt w^{-a}$ or $\zt y w^{-a}$.  

Motivated by the form of the spectral curve of the integrable system, as discussed above, we choose the to put the stuck 5 brane at $S^+_\infty$.  This means that the $A_{a\ell}$ and $E_{a\ell}$ coefficients are determined by requiring that all second- and higher-order poles along $S^\pm_\infty$ and the simple poles along $S^-_\infty$ cancel in the $\zt$ variables.  Only a simple pole along $S^+_\infty$ is allowed, corresponding to the stuck brane.

\subsubsection{The rank-1 SW curve}

Specializing to rank $N=1$, there is no $M$ deformation, and the M5 brane curve \eqref{Q0M5crv} becomes simply
\begin{align}\label{rk1SWcrv}
0=z +A_{01} .
\end{align}
We can use this to eliminate $z $ in the \eqref{surfwp} to give the an elliptic curve in Weierstrass form for the SW curve.  We recall here for later convenience the expressions for the $Q_0$ surface and 1-form written in the $w=1$ patch coordinates,
\begin{align}\label{SW1form}
y^2 &= (z  + \m_0^2 x) \til P + \til Q, &
\l &= \frac{y dx}{\til P} ,
\end{align}
where
\begin{align}\label{}
\til P &:= \prod_{i=1}^3 (x-e_i) ,&
\til Q & := \sum_{j=1}^3 \m_j^2 \e_j \prod_{k\neq j} (x - e_k) .
\end{align}

\subsubsection{The rank-2 SW curve}

At rank $N=2$ the coefficients in the general M5 brane curve \eqref{QMcur} are determined by the procedure described below equation \eqref{case2Spcrv}.  For $N=2$ the highest power of $y$ appearing in \eqref{case2Spcrv} is 2, and only a single iteration of using the $Q_0$ surface equation to reduce the power of $y$ is needed.  As a result the constraints on the coefficients are not overly complicated, though it is still useful to use a computer algebra system to solve the constraints.  The result is that the M5 brane curve is (written in the $w=1$ patch coordinates)
\begin{align}\label{rk2SWcrv}
0=z ^2+A_{01} z  + A_{20} 
- 4 M^2 z x
- 8 M^2 \m_0 (y+\m_0 x^2) .
\end{align}
The intersection of \eqref{rk2SWcrv} with the $Q_0$ surface \eqref{SW1form} and the restriction of the one-form to this intersection then give a genus-2 SW curve and associated meromorphic 1-form.

\section{Matching spectral curve to M5 brane curve }
\label{section: BCtoM5_1}

The Coulomb branch of the $\USp(2N)$ $N_f=4$ theory is isomorphic as a complex space (though not as a metric space) to $\C^N$ with coordinates given by the gauge invariant vacuum expectation values $u_i := \tr(\wedge^{2i}\Phi)$, $i = 1, 2, \ldots, N$ which have scaling dimensions $2, 4, \ldots, 2N$ at the conformal point.  The Coulomb branch of the massless theory has the same complex structure as the classical moduli space.  At a generic point on the Coulomb branch of the massless theory, the adjoint vev can be diagonalized, $\Phi = \mathrm{diag}(\pm \phi_1, \pm \phi_2, \cdots, \pm \phi_N)$, in which case $u_i = e_i(\phi_1^2, \phi_2^2, \cdots, \phi_N^2)$, $i =1, 2, \ldots, N$, where $e_i$ is the $i$-th elementary symmetric polynomial.  As long as the antisymmetric mass vanishes, the matrix of $\U(1)^N$ complex gauge couplings is diagonal, $\t_{ij} = \d_{ij} \t(\phi_i^2)$.  

In the case when all the masses vanish, $\t(\phi_i^2) = \t$, i.e., has the same constant value.  We thus have the same abelian variety with period matrix $\tau_{ij} = \delta_{ij} \tau$ at all points on the Coulomb branch except the origin. The singular fiber above the origin is given by the orbifold $T^{2N}/G(2,1,N) \simeq \mathbb{C}^N / ([\Z + \tau \Z)^N \rtimes G(2,1,N)]$.  Then the total space of Coulomb branch is identical to the phase space of the Inozemtsev system with zero couplings.  

Thus for vanishing masses the field theory Coulomb branch geometry is correctly described by the Inozemtsev system.   
In the remainder of this section we present parameter and variable identifications for the rank $N=1$ and $N=2$ cases, showing that the M5 brane SW curve and 1-form and the spectral curve and 1-form of the Inozemtsev system coincide for non-vanishing masses (deformation parameters).  
We stop at $N=2$ because the matching of parameters becomes increasingly complicated for larger values of $N$.

\subsection{The $N=1$ case}

Recall that the $N=1$ spectral curve is given by \eqref{BC1curve}, and the one-form by $\l=k d\alpha$.  Introduce coordinates $(x,y)$ related to $(k,\alpha)$ by
\begin{align}\label{kz2xy}
x & =  \wp(\alpha), &
y & =  \frac14 \wp'(\alpha) k,
\end{align}
where the prime means derivative with respect to $\alpha$.  These definitions were motivated in \eqref{xydef} by the pole structure of the spectral curve.   We then find, using the Weierstrass $\wp$-function identities
\begin{align}\label{wpidents}
(\wp'(\alpha))^2 &= 4\prod_{i=1}^3 (\wp(\alpha)-e_i), &
\wp(\alpha+\om_i) &= e_i +\frac{\prod_{j\neq i}^3 (e_i-e_j)}{\wp(\alpha)-e_i} ,
\end{align}
where
\begin{align}\label{}
e_i := \wp(\om_i), \quad i=1,2,3 ,
\end{align}
that the spectral curve and one-form become
\begin{align}\label{}
y^2  &=  \tfrac14 (h_2 + \g) \prod^3_{i=1} (x-e_i)  
+ \tfrac14 (g_0^\vee)^2 x \prod^3_{i=1} (x-e_i) 
+ \tfrac14 \sum^3_{i=1} (g_i^\vee)^2  \prod^3_{j\neq i} (x-e_i) (e_i-e_j) ,
\nn\\
kd\alpha 
&=  \frac{y dx}{ \prod^3_{i=1} (x-e_i) } ,
\end{align}
where $\g : = \sum^3_{i=1} (g_i^\vee)^2 e_i$.  These are easily seen to coincide with the SW curve and 1-form given in \eqref{rk1SWcrv} and \eqref{SW1form} with the parameter identifications
\begin{align}\label{}
\mu_i^2 &= \frac14 (g_i^\vee)^2 , &
A_{01} &= -\frac14 (h_2 + \g) .
\end{align}

\subsection{The $N=2$ case}
\label{section: BCtoM5_2}

Recall that the $BC_2$ spectral curve is given by \eqref{BC2crv}.  With the same change of variables \eqref{kz2xy} as in the $BC_1$ case, which matched the 1-forms, the $BC_2$ curve becomes
\begin{align}\label{}
(k^2 - u^\vee)^2 - h_2 (k^2 - u^\lor) + h_4 
- 4g^2 \left( x (k^2 - u^\vee) + 4g^\lor_0 y  
+ 2(g^\vee_0)^2 x^2  
+ x \g  \right) = 0.
\end{align}
Recall that $u^\vee := \sum_{r=0}^3 (g_r^\vee)^2 \wp(\alpha + \om_r)$ and $ \g := \sum_{r=1}^3 (g^\lor_r)^2 e_r$.  Then with the parameter identifications
\begin{align}\label{matchn=2}
\m_0 & =   \frac{1}{2} g_0^\vee, &
\m_i^2 &= \frac{1}{4} (g_i^\vee)^2 \ \text{  for }\ i \in \{1,2,3\}, &  
M^2 & = \frac{1}{4} g^2,
\nn\\
A_{01} & = - \frac{1}{4} (h_2 + 2  \g), &
A_{02} & = \frac{1}{16} (h_4 + \g h_2 + \g^2),
\nn\\
z & = \frac14 (k^2 - u^\vee + \g),
\end{align}
and using the Weierstrass identities \eqref{wpidents}, we find the spectral curve becomes the pair of equations
\begin{align}
y^2  &= (z + \m_0^2 x ) \til P + \til Q, \nn \\
0 &= z^2 + z \left(A_{01}- 4M^2 x \right) 
+ \left( A_{02} - 8 M^2 \m_0^2 x^2 \right) 
-8 M^2 \m_0 y,
\end{align}
which coincides with the M5 brane curve \ref{rk2SWcrv} and background surface \ref{SW1form}.   Note that the definition of $z$ (up to a constant shift) was already motivated in \eqref{u-vee} by the pole structure of the spectral curve.

\section*{Acknowledgement}
We would like to thank Mario Martone, Joseph Minahan, Yiwen Pan, Oliver Schlotterer for helpful discussions. YL is grateful to Yuji Tachikawa for pointing out the paper \cite{GR12} and for encouragement. The collaboration between OC and YL started during the workshop ``Elliptic integrable systems, special functions, and quantum field theory" which took place in June, 2019 at Nordita, Stockholm. OC is grateful to the organizers for the invitation, and OC and YL would like to thank Nordita for the hospitality and stimulating environment. The work of PCA is supported in part by DOE grant SC0011784. The work of YL is supported by the Knut and Alice Wallenberg Foundation under grant Dnr KAW 2015.0083.

\appendix
\section{Appendix}

\subsection{Elliptic functions and identities}\label{appx:ell}
We use the following functions 
\begin{equation}\label{sir}
	\sigma_\alpha^r(x) = \frac{\vartheta_{r+1}(x- \alpha) \vartheta_1 ' (0)}{\vartheta_{r+1}(x) \vartheta_1 (- \alpha)}\,,\quad r=0, 1, 2, 3\,,
\end{equation}
where $\vartheta_{1,2,3,4}(x|\tau)$ are the Jacobi theta functions. A summary of their main properties can be found in \cite{KH97}; in particular, we have
\begin{equation}
	\sigma_\alpha^r(x + \omega ) = e^{2 \pi i \alpha \partial_\tau \omega} \sigma_\alpha^r(x) \quad\text{for}\ \omega \in \Z + \tau \Z\,.
\end{equation}
(Here we use the shorthand notation $\partial_\tau(a+b\tau)=b$.) We also denote $\sigma_\alpha^0(x)$ simply by $\sigma_\alpha(x)$, that is,
\begin{equation}
	\sigma_\alpha(x) = \frac{\vartheta_{1}(x- \alpha) \vartheta_1 ' (0)}{\vartheta_{1}(x) \vartheta_1 (- \alpha)}\,.
\end{equation}
The functions \eqref{sir} are related to each other by translations by the half-periods $(\omega_0, \omega_1, \omega_2, \omega_3)=(0, \frac12, \frac{1+\tau}{2}, \frac{\tau}{2})$:
\begin{equation}
	\sigma_\alpha^r(x) = e^{2\pi i \alpha \partial_\tau \omega_r } \sigma_\alpha(x - \omega_r).
\end{equation}
For given coupling parameters $g_{0,1,2,3}$, we further define
\begin{equation}
    v_\alpha(x) =v_{\alpha}(x; g_0, g_1, g_2, g_3)=\sum_{r=0}^{3} g_r \sigma_{2\alpha}^r(x) .
\end{equation}
Note the properties
\begin{align}\label{vv}
\sigma_{-\alpha}(-x) & =-\sigma_{\alpha}(x)\,,\quad v_{-\alpha}(-x)=-v_\alpha(x)\,,
\end{align}
and the following identities:
\begin{align}
\label{adds}
\sigma_\alpha(-x) \sigma_\alpha(x) & = \wp(\alpha) - \wp(x),\\
\label{addv}
v_\alpha(-x) v_\alpha(x)  & = \sum_{r=0}^3 \Big( (g_r^\vee)^2 \wp( \alpha + \omega_r) -  (g_r)^2 \wp( x + \omega_r) \Big)\,,
\end{align}
where $g^\vee_i$ are the dual parameters \eqref{ghat}. Using the notation \eqref{u}, \eqref{u-vee}, the last relation can be written as $v_\alpha(-x)v_\alpha(x)=u^\vee(\alpha)-u(x)$.

Another useful property of $v_\alpha(x)$ is the following duality:
\begin{equation}\label{vsym}
v_\alpha(x; g_0, g_1, g_2, g_3)=v_{-x}(-\alpha;  g^\vee_0, g^\vee_1, g^\vee_2, g^\vee_3)=-v_x(\alpha;  g^\vee_0, g^\vee_1, g^\vee_2, g^\vee_3)\,.
\end{equation}
This can be checked by comparing translation properties and residues in the $x$-variable. 

Finally, let us state how $\sigma_\alpha(x)$ and $v_\alpha(x)$ behave under action of $\gamma\in\SL(2, \Z)$. We will use the group homomorphism $\pi$ from $\SL(2, \C)$ to the permutation group $S_3$ defined on the generators as follows: 
\begin{equation}\label{hom}
\pi\,:\,\SL(2, \C)\to S_3\,,\quad \gamma\mapsto \pi_\gamma\,,\quad \left(\bsm 1 & 1\\ 0 & 1 \esm\right)\mapsto s_{23}\,,\quad \left(\bsm 0 & 1\\ -1 & 0 \esm\right)\mapsto s_{13}\,.
\end{equation}
Note that the kernel of $\pi$ is the principal congruence subgroup  $\G(2)\subset \SL(2, \C)$.

Take $\g =\left(\bsm a & b\\ c & d \esm\right) \in \SL(2,\Z)$ and define $\tau'$, $\alpha'$, $x'$, $g'_i$ in the following way:
\begin{align}
\t' & = \frac{a\t +b}{c\t +d}\,,\quad \alpha'  = (c\t+d)^{-1}\alpha\,,\quad x' = (c\t+d)^{-1} x\,,\\
g_0' & =g_0\,,\quad g'_r = g_{\pi_{\g}(r)}\quad \text{for}\ r=1,2,3\,.
\end{align}
With this notation, we have:  
\begin{align}
\s_{\alpha'}(x'|\t') & = (c\t +d) \exp\left(-\frac{2\pi ic}{c\t+d} \alpha x\right) \s_\alpha(x|\t)\,,\\
\s^{\pi_\gamma(r)}_{\alpha'}(x'|\t') & = (c\t+d) \exp\left(-\frac{2\pi ic}{c\t+d} \alpha x\right) \s^r_\alpha(x|\t)\,,\quad r=1,2,3\,.
\end{align}
These transformations can be deduced easily using the modular transformations of Jacobi theta functions. As a corollary, 
\begin{align}
v_{\alpha'}(x'; g_0', g_1', g_2', g_3'|\t') = (c\t+d) \exp\left(-\frac{4\pi ic}{c\t+d} \alpha x\right) v_\alpha(x; g_0, g_1, g_2, g_3|\t)\,.
\end{align}

\subsection{Calculating the $N=2$ spectral curve}\label{appB}

The $N=2$ spectral curve is defined by the characteristic polynomial 
\begin{equation}
\det( L- k \mathrm{Id}) = k^4+a_1k^3+a_2k^2+a_3k+a_4
\end{equation}
of the Lax matrix \eqref{n2mat}. By direct calculation,
\begin{align}
a_1 & =  0  \nn \\
a_2 & =   - \Bigl( p_1^2 -p_2^2 
+ 2 g^2 \bigl( \sigma_\alpha(-q_{12})  \sigma_\alpha(q_{12}) +  \sigma_\alpha(-q^+_{12}) \sigma_\alpha(q^{+}_{12}) \bigr) 
\nn \\
&  \quad + v_\alpha(-q_1) v_\alpha ( q_1) + v_\alpha(-q_2) v_\alpha (q_2) \Bigr)
\nn \\
a_3 & =  -2 g^2  \Big( v_\alpha ( q_2 ) \sigma_\alpha ( q_{12} ) \sigma_\alpha ( -q^{+}_{12} )  + v_\alpha ( q_1 ) \sigma_\alpha ( -q_{12} ) \sigma_\alpha ( -q^{+}_{12} ) 
\nn \\
& \quad  + v_\alpha (- q_1 ) \sigma_\alpha ( q_{12} ) \sigma_\alpha ( q^{+}_{12} ) + v_\alpha (- q_2 ) \sigma_\alpha ( -q_{12} ) \sigma_\alpha ( q^{+}_{12} )  \Big) 
\nn \\
a_4& =   p_1^2 p_2^2 +   v_\alpha (- q_2 ) v_\alpha ( q_2 ) p_1^2 + v_\alpha (- q_1 ) v_\alpha ( q_1 ) p_2^2 
\nn \\
&  \quad + 2 g^2 \Big( \sigma_\alpha ( -q^{+}_{12} ) \sigma_\alpha ( q^{+}_{12} ) - \sigma_\alpha ( -q_{12} ) \sigma_\alpha ( q_{12} )  \Big) p_1 p_2 
\nn  \\
&  \quad +v_\alpha (- q_1 ) v_\alpha ( q_1 ) v_\alpha (- q_2 ) v_\alpha ( q_2 ) 
\nn \\
&  \quad -g^2 \Big( v_\alpha ( q_1 ) v_\alpha ( q_2 ) \sigma_\alpha ( -q^{+}_{12} )^2 + v_\alpha (- q_1 ) v_\alpha ( q_2 ) \sigma_\alpha ( q_{12} )^2 
\nn \\
&  \quad + v_\alpha ( q_1 ) v_\alpha (- q_2 ) \sigma_\alpha ( -q_{12} )^2 + v_\alpha (- q_1 ) v_\alpha (- q_2 ) \sigma_\alpha ( q^{+}_{12} )^2 \Big) 
\nn\\
& \quad + g^4 \Big(  \sigma_\alpha ( -q_{12} ) \sigma_\alpha ( q_{12} ) - \sigma_\alpha ( -q^{+}_{12} ) \sigma_\alpha ( q^{+}_{12} ) \Big)^2
\end{align}
where we have used the abbreviations $q_{ij} = q_i - q_j$ and $q_{ij}^+ = q_i + q_j$.

Using \eqref{adds} and \eqref{addv}, we easily find that
\begin{align}
a_2 & = - \Big( h_2 + 4 g^2 \wp(\alpha) + 2 \sum_{r=0}^3 (g_r^\vee)^2 \wp(\alpha + \omega_r) \Big) 
= - (h_2 + 4 g^2 \wp(\alpha) + 2 u^\vee(\alpha))\,,
\end{align}
where 
\begin{equation}\label{ih2}
h_2 =p_1^2+p_2^2 - u(q_1)-u(q_2) -2g^2\left( \wp(q_{12})  + \wp(q_{12}^+)\right)\,. 
\end{equation}

To calculate $a_3$, we first note that it is elliptic in $q_{1,2}$ with possible first order poles along the mirrors $q_{i} =0$ for $i = 1,2$ and $q_{1} \pm q_{2} = 0$.  However, it is symmetric under interchanging $q_1, q_2$ and changing their signs arbitrarily. Hence, $a_3$ cannot have a first order pole along any mirror, thus it is regular elliptic, i.e. constant independent of $q_1$, $q_2$. After than we can evaluate $a_2$ at convenient values of $q_1, q_2$. The result is 
\begin{align}\label{}
a_3 & = - 2 g^2 \Big( \sum_{i=0}^3 g_i \Big) \wp'(\alpha) 
= - 4 g^2 g_0^\vee \wp'(\alpha) .
\end{align}

It remains to deal with $a_4$. By using \eqref{adds} and \eqref{addv} repeatedly, we rearrange it into
\begin{align}\label{a44}
a_4 
& = 
 \left( \sum_{r=0}^3 (g_r^\vee)^2 \wp(\alpha + \omega_r) \right) h_2 
+ \left(\sum_{r=0}^3 (g_r^\vee)^2 \wp(\alpha + \omega_r) \right)^2 
\nn\\
& \qquad + \left(p_1 p_2 +  g^2\wp(q_{12}) - g^2\wp(q_{12}^+) \right)^2 \nn\\
& \qquad \mbox{}- u(q_1) p_2^2  - u(q_2) p_1^2  +  u(q_1) u(q_2)
- g^2b 
\end{align}
where we have introduced 
\begin{align}\label{}
b & =  v_\alpha ( q_1 ) v_\alpha ( q_2 ) \sigma_\alpha ( -q^{+}_{12} )^2 + v_\alpha (- q_1 ) v_\alpha ( q_2 ) \sigma_\alpha ( q_{12} )^2 
\nn\\
& \ \ \qquad + v_\alpha ( q_1 ) v_\alpha (- q_2 ) \sigma_\alpha ( -q_{12} )^2 + v_\alpha (- q_1 ) v_\alpha (- q_2 ) \sigma_\alpha ( q^{+}_{12} )^2\,.
\end{align}
Calculating $b$ is more involved, so we just give a sketch. As the first step, we analyse the $2$nd order poles in $q_1, q_2$ and find that  
the following expression agrees with $b$ up to an extra term $c$ having first order poles only:
\begin{align}\label{bb}
	b & = ( 2 u^\vee(\alpha) -  u (q_1) - u(q_2) ) (\wp(q_{12}) + \wp(q_{12}^+))
	+ \sum_{r=1}^{3} 2 g_r^2 \wp(q_1 + \omega_{r}) \wp(q_2 + \omega_{r}) +c\,.
\end{align}
Using the symmetry arguments once more, we conclude that $c$ must be regular, i.e. it is just a function of $\alpha$. In addition, we know that $c=c(\alpha)$ is even elliptic. It is also easy to check that $c(\alpha)$ has a 4th order pole at $\alpha=0$ and 2nd order poles at $\alpha = \omega_{1,2,3}$. To determine $c(\alpha)$ from that, we analyse the Laurent expansion of $b$ in $\alpha$ near $\alpha=0$ and $\alpha=\omega_{1,2,3}$. We skip the details and just give the answer:
\begin{align} \label{eq:falpha}
	c & = 2 \left[ 4 (g_0^\vee)^2 \wp(\alpha)^2 - 2 \wp(\alpha) \left( u^\vee(\alpha) - \sum_{i=1}^{3} (g^\vee_r)^2 e_r \right) \right]+d\,,
	\end{align}
	up to a possible constant $d$ which may depend on $g_i$ and $e_i=\wp(\omega_i)$, but not on $\alpha$. 

Backward substitution of \eqref{eq:falpha} and \eqref{bb} into \eqref{a44} gives the answer for $a_4$, after which all that remains is to rearrange the terms based on the form of the quartic Hamiltonian $h_4$ \eqref{h4}. The constant $d$ in \eqref{eq:falpha} can always be absorbed into $h_4$, so can be ignored. 

\bibliography{Inozemtsev_SW_bib} 
\bibliographystyle{JHEP}

\end{document}